\documentclass[letterpaper,journal]{IEEEtran}
\usepackage{amsmath,amsfonts}
\usepackage{algorithmic}
\usepackage{algorithm}
\usepackage{array}
\usepackage[caption=false,font=normalsize,labelfont=sf,textfont=sf]{subfig}
\usepackage{textcomp}
\usepackage{stfloats}
\usepackage{url}
\usepackage{verbatim}
\usepackage{graphicx}
\usepackage{cite}
\usepackage{multirow}
\usepackage{capt-of}
\usepackage[table]{xcolor}

\makeatletter
\begingroup
\normalsize
\Huge
\endgroup
\makeatother

\begin{document}

\title{Beyond Semantic Prompts: Unified Watermarking and Access Control for Text-to-Image Diffusion Models}

\author{Fangming Shi, Li Li, Guorui Feng, Xinpeng Zhang
\thanks{This work was supported in part by the National Natural Science Foundation of China under the grant No. 62302286. \textit{(Corresponding author: Li Li.})}%
\thanks{Fangming Shi is with the School of Communication and Information Engineering, Shanghai University, Shanghai 200444, China (e-mail: shifangming0823@shu.edu.cn).}
\thanks{Li Li and Guorui Feng are with the School of Communication and Information Engineering, Shanghai University, Shanghai 200444, China, and also with the National Key Laboratory of Maritime Unmanned Swarm Intelligence (e-mail: \{llichn, grfeng\}@shu.edu.cn).}
\thanks{Xinpeng Zhang is with the School of Computer Engineering and Science, Shanghai University, Shanghai 200444, China, and also with the National Key Laboratory of Maritime Unmanned Swarm Intelligence (e-mail: xzhang@shu.edu.cn).}}

\markboth{}%
{}


\maketitle

\begin{abstract}
	Text-to-image diffusion services require both traceability of generated content and differentiated access to generation capabilities. However, existing watermarking methods commonly expose watermark control through mechanisms beyond the text-prompt interface, while credential-based access control does not inherently enforce distinct output quality levels. We propose a unified framework that enables customizable multi-bit watermark embedding and tiered access control through a single text prompt. A unified prompt represents watermark messages using designated words and associates protocol structures with different access states. The framework adopts a two-stage training strategy. First, watermark payload partitioning represents messages as designated watermark words, while vocabulary extension assigns each protocol word a dedicated token with a learnable input embedding. Guided by a frozen reference model and a pretrained watermark codec, joint adaptation of the U-Net and protocol-token embeddings transfers message-dependent watermark patterns into the generation process. Second, a prompt-structure recognition network controls feature modulation within the variational autoencoder decoder, which is trained to produce clear watermarked images, blurred previews, or black outputs according to the access state. Experiments demonstrate reliable watermark recovery, robustness to common image-processing operations and evaluated model modifications, and consistent performance across diverse generation settings. The controlled decoder also achieves distinct output behaviors under restricted access while closely preserving reference decoding fidelity and watermark recoverability under full access. Our open-source code and model weights will be public after acceptance.
\end{abstract}


\section{Introduction}
\IEEEPARstart{T}{}ext-to-image diffusion models have made high-quality image generation increasingly accessible through natural-language prompts. Beyond generating images, commercial generation services must balance user experience with revenue generation. For example, providers may offer free trials with low-quality previews while reserving full-quality generation for paying users, thereby attracting potential customers and encouraging subscription upgrades. This business model requires generation systems to deliver outputs with different quality and usability levels according to users' authorization states. Meanwhile, the widespread dissemination of AI-generated content has heightened the need for copyright protection and provenance traceability. 
Service providers need reliable mechanisms to associate generated images with the platform users that requested their generation, supporting provenance tracing and and user-level attribution of generated content.
These requirements raise a practical challenge for text-to-image generation services: how can a unified generation pipeline regulate the form of delivered outputs while preserving traceable copyright information in the generated images?

\begin{figure}[!t]
	\centering 
	\includegraphics[
	width=\columnwidth]{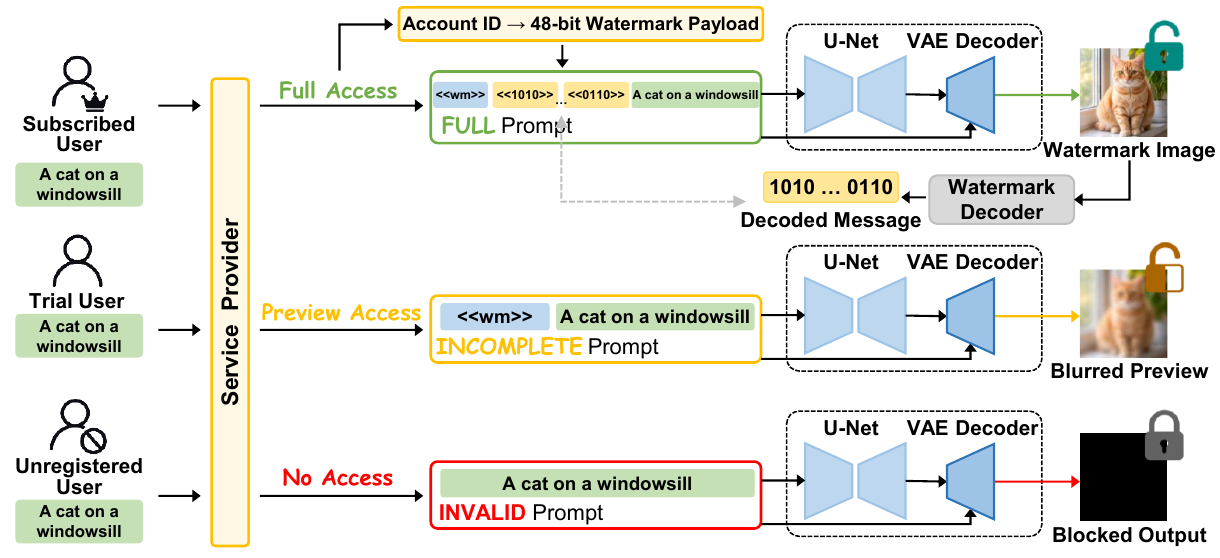}
	\caption{Overview of customizable watermarking and tiered access control through a unified prompt protocol. Users submit semantic descriptions, and the service provider assembles model-facing prompts according to current subscription status. FULL prompts produce high-quality images carrying the specified watermark messages; INCOMPLETE prompts yield blurred previews; and INVALID prompts produce black outputs that suppress visible content.}
	\label{fig:introduction}
\end{figure}

\IEEEpubidadjcol

Access control and digital watermarking address two important requirements of commercial text-to-image generation services. Access control determines the availability and quality of generated outputs according to authorization policies, whereas watermarking embeds recoverable information into images to support ownership attribution and provenance tracing~\cite{luo2025digital}. Despite their complementary roles, existing approaches typically implement these capabilities through separate mechanisms. For access control, existing methods regulate generation through modifications to internal conditioning mechanisms~\cite{bui2025hiding} or noise distributions~\cite{tanaka2025access}, which may affect the generation process and compromise fidelity under authorized use. Key-based methods~\cite{gai2025pcdiff} provide another means of distinguishing valid from invalid credentials, but credential verification alone does not necessarily support explicitly differentiated output behaviors, such as providing blurred previews to trial users and suppressing usable content for unauthorized requests. For watermarking, existing diffusion-based methods embed provenance information through initial noise~\cite{wen2023treering,yang2024gaussian}, model parameters~\cite{feng2024aqualora}, or dedicated conditioning mechanisms~\cite{min2025wadiff}, enabling generated images to carry recoverable ownership or source information. However, existing approaches lack a unified mechanism to coordinate access control and watermarking during image generation. As a result, output access and fidelity are regulated independently of provenance embedding, limiting the ability of commercial services to provide differentiated outputs while ensuring consistent watermark-based traceability.

Building on this motivation, we propose a unified prompt-protocol framework that jointly encodes watermark messages and output-control instructions into text prompts, integrating provenance tracing and tiered access control within a single generation pipeline. 
As illustrated in Fig.~\ref{fig:introduction}, users submit semantic descriptions to the service provider. For each request, the service provider determines the user's current access privileges, encodes the platform-assigned user identifier as a watermark payload when full access is granted, and constructs the corresponding model-facing prompt. The diffusion model then produces a high-quality watermarked image, a blurred preview, or a content-suppressed output corresponding to the prompt.
Realizing this framework involves two key challenges: i) \textbf{How can watermark information specified by designated protocol words be reliably embedded into generated images while ensuring recoverability?} Unlike semantic words, watermark words must convey specific payload information and establish a reliable correspondence between textual representations and recoverable image-embedded patterns. ii) \textbf{How can compositional prompt protocols reliably control generation behaviors across different access levels?} This requires translating protocol compositions into consistent output behaviors while preserving the intended fidelity distinctions.  To address these challenges, we develop a prompt-conditioned watermark embedding method that associates designated protocol words with recoverable watermark patterns, and a protocol-conditioned access control mechanism that interprets prompt composition rules to regulate output fidelity.

Our main contributions are summarized as follows:
\begin{itemize}
	\item We propose a unified prompt-protocol framework that integrates platform-managed multi-bit watermark embedding with tiered output control for text-to-image diffusion services. Unlike approaches that treat watermarking and output control as independent operations, our framework jointly encodes watermark messages and output fidelity instructions into a shared prompt protocol, enabling both functions to be realized within a single generation process.
	
	\item We develop a prompt-conditioned watermark embedding method that jointly optimizes the U-Net and protocol-token embeddings under the guidance of a reference model and prompts containing designated protocol words. By encouraging the model to preserve pretrained message-dependent watermark patterns during denoising and associate them with the corresponding protocol words, the proposed method enables prompt-specified watermark generation while maintaining semantic generation capabilities.
	
	\item We design a protocol-conditioned output-control mechanism based on prompt-structure and decoder-feature modulation. With explicit supervision for three output states, the mechanism produces high-fidelity watermarked images, blurred previews, and content-suppressed outputs according to the specified protocol.
	
\end{itemize}

\section{Related Work}

\subsection{Text-to-Image Diffusion Models}

Advances in diffusion modeling, including DDPM~\cite{ho2020ddpm}, DDIM~\cite{song2021ddim}, and latent diffusion models (LDMs)~\cite{rombach2022ldm}, have enabled high-quality image synthesis with increasingly efficient sampling and computation. Built on the latent diffusion framework, Stable Diffusion~\cite{rombach2022sd15} has gained widespread adoption among researchers and digital artists owing to its high-quality image generation ability and publicly available code and model weights.

In the U-Net-based architecture considered in this work, a tokenizer converts the input text prompt into token IDs, a text encoder~\cite{radford2021clip} maps these IDs to contextualized token representations, a conditional U-Net iteratively denoises the latent representation, and a variational autoencoder (VAE)~\cite{kingma2014vae} decoder reconstructs the final image. Our work instead investigates how the prompt can additionally convey watermark messages and control decoding behavior, coordinating traceability watermarking in the denoising network with generation-quality control in the VAE decoder.

\subsection{Watermarking for Text-to-Image Diffusion Models}

The widespread availability of text-to-image models creates a need to identify synthetic content and trace its generator. Image watermarking addresses this need by embedding recoverable signals that support source attribution and the investigation of unauthorized use. To improve watermark robustness, numerous methods embed traceability watermarks directly during image generation. 
Two prominent categories~\cite{luo2025digital} are initial-noise-based watermarking and fine-tuning-based watermarking. Initial-noise-based methods~\cite{wen2023treering,yang2024gaussian,luo2026flexible,li2025stdm} encode watermark information in the starting noise of the reverse denoising process, while fine-tuning-based methods~\cite{fernandez2023stable,feng2024aqualora,min2025wadiff} adapt the denoising network or image decoder to produce watermarked outputs. 

Despite their effectiveness, these approaches generally control watermark through mechanisms separate from the text prompt, such as noise construction, signature-specific decoder parameters, LoRA scaling matrices, or dedicated watermark-conditioning inputs. Consequently, deploying these methods or changing their watermark payloads can require access to generation-pipeline internals or model-level configurations. Such control mechanisms can introduce additional deployment complexity and computational overhead for service providers, compromise the consistency of model behavior, and create new attack surfaces.


Our approach represents variable multi-bit watermark messages as designated words within the model-facing prompt. Once the adapted model is deployed, the service provider can encode the requesting user's platform-assigned identifier and coordinate watermark embedding with output control through server-side prompt assembly, without manipulating model parameters or generation-pipeline internals.

\subsection{Access Control for Text-to-Image Diffusion Models}
Access control for text-to-image diffusion models aims to regulate the use of generative capabilities according to users' authorization levels, enabling service providers to offer differentiated generation services while protecting their proprietary models and generation resources. Unlike conventional access control, which primarily governs whether a user can invoke a service, access control in generative models can be enforced directly within the generation pipeline, regulating not only whether an output can be obtained but also the quality and completeness of the generated content. Existing approaches enforce access restrictions at different stages of the generation pipeline, including semantic conditioning, sampling initialization, and latent decoding. At the conditioning stage, Bui et al.~\cite{bui2025hiding} introduce learnable prompts to suppress selected concepts and enable their recovery with a secret key, thereby restricting the semantic content that can be generated. At the sampling stage, Tanaka and Nakamura~\cite{tanaka2025access} regulate generation quality by replacing the standard Gaussian initialization with a key-dependent noise distribution, making high-quality generation dependent on the corresponding key. At the decoding stage, PCDiff~\cite{gai2025pcdiff} introduces key-conditioned fuser layers and additional fine-tuning layers into the VAE decoder to regulate the quality of reconstructed images. 

Despite these advances, existing methods do not explicitly integrate access control with watermark-based provenance tracing. Although PCDiff~\cite{gai2025pcdiff} claims compatibility with existing watermarking techniques, their integration remains challenging. First, both access control and several watermarking methods modify the VAE decoder~\cite{fernandez2023stable,lin2026ewlora}, potentially introducing functional interference when combined. Second, loosely coupled implementations may expose additional attack surfaces, allowing adversaries to disrupt watermark embedding or circumvent access restrictions. These limitations motivate a unified design that jointly coordinates watermark embedding and access control.


To address these challenges, we couple watermark specification and model access control within a unified text-prompt protocol. The U-Net embeds the prompt-specified watermark message during image generation, while an additional module parses the prompt structure to guide the VAE decoder toward the output quality associated with the specified access state. Both functions are jointly optimized within the unified framework, enabling coordinated watermark embedding and access control while minimizing interference between their respective mechanisms.

\section{Threat Model}

\subsection{User Roles and Capabilities}

We consider a commercial text-to-image generation service that provides tiered access and embeds provenance watermarks in generated images. This setting involves a service provider, users with different subscription status and corresponding access levels, and malicious third parties who attempt to remove watermarks from publicly shared images.

\textbf{Service provider.} It offers online image generation services through an API. For each request, it determines the requesting user's identity and current subscription status, and constructs the model-facing prompt from the submitted semantic description. For full-access requests, it encodes the platform-assigned user identifier as the watermark payload. 

\textbf{Users.} Users submit semantic descriptions through the service prompt interface. Their access levels are determined by the provider's current subscription status.

\textbf{1) Unregistered users.} The provider assigns requests from unregistered users to the no-access state and forwards the semantic description without protocol words as an INVALID prompt. The model produces a uniformly black image.

\textbf{2) Trial users.} The provider assigns eligible trial users to the preview state and constructs an INCOMPLETE prompt, by prepending only the designated trigger word to the semantic description. The model produces a blurred preview that retains coarse semantic content while restricting access to full-quality outputs.

\textbf{3) Subscribed users.} For users with an active subscription granting full access, the provider constructs a FULL prompt by combining the trigger word, the watermark-word sequence encoding the user's platform-assigned identifier, and the semantic description. The model produces a high-quality image carrying a recoverable watermark associated with the requesting user.

\textbf{Malicious third parties}
They may obtain high-quality, watermarked images published by creators on public platforms. The adversary seeks to remove the embedded watermark while preserving visual quality, with the aim of evading provenance tracing and falsely claiming ownership of the images.

\subsection{Defense Objective}
\subsubsection{Base Objective}
Our defense has two complementary objectives: enforcing tiered model access and preserving the effectiveness of prompt-specified watermarks. For access control, the model should preserve high-quality generation under full access, while producing outputs consistent with the prescribed access state, with sufficiently distinct visual quality and usability across access levels. For watermarking, the embedded message should be reliably recoverable while introducing minimal perceptible distortion. Watermark recovery should also remain robust to the image-processing operations and model modifications considered in this work.

\subsubsection{Threats and Attacks}
\textbf{Access-control attacks.} We consider users without current full-access privileges, including trial users, unregistered users, and users whose subscriptions have expired. These adversaries may attempt to obtain high-quality generation services without satisfying the required authorization conditions. They may attempt to bypass access restrictions by applying post-processing techniques, including image super-resolution, denoising, and blind restoration, to restricted previews in an attempt to recover visually usable, high-quality images without legitimate full access.

\textbf{Watermark attacks}. We consider two categories of attacks against watermark integrity and recoverability:
i) \textbf{Image-level attacks}. Generated images may undergo routine distortions during storage, transmission, and subsequent use, such as additive noise, cropping, and compression. These transformations may unintentionally affect watermark detection. More importantly, an adversary who intends to appropriate valuable generated images may deliberately remove or obscure the embedded watermark while retaining as much visual quality as possible. 
ii) \textbf{Model-level attacks}. We also consider modifications applied by service providers to their deployed models. The service provider has full access to the deployed model and may modify it for computational efficiency or additional functionality, such as supporting stylized image generation. Although these operations do not intentionally remove the watermark, they may alter watermark-bearing representations and impair subsequent watermark recovery. 


\section{Methodology}
\subsection{Problem Formulation}

Our objective is to jointly enforce watermark embedding and output quality through the input prompt. Let $c$ denote a semantic caption containing no protocol words, and let $\mathbf{m}\in\{0,1\}^{48}$ denote a watermark message. An input prompt $p$ combines $c$ with optional protocol words that encode the watermark payload and determine the access state. We denote the state associated with its protocol structure by
\begin{equation}
	s(p)\in\{\mathrm{FULL},\mathrm{INCOMPLETE},\mathrm{INVALID}\},
\end{equation}
corresponding to full, preview, and no access, respectively. 

We formulate the generation process as
\begin{equation}
	\mathbf{z}=F_{\theta}(p,\boldsymbol{\epsilon}),
	\qquad
	\mathbf{x}=D_{\phi}(\mathbf{z},p),
\end{equation}
where $\boldsymbol{\epsilon}\sim\mathcal{N}(\mathbf{0},\mathbf{I})$ is the initial noise, $F_{\theta}$ denotes the text-conditioned latent generation process comprising the text encoder and U-Net, and $D_{\phi}$ denotes the VAE decoder augmented with prompt-dependent control. 

The desired behavior comprises two requirements. First, a FULL prompt should produce a high-quality image consistent with $c$, from which the specified watermark message can be reliably recovered. Let $D_s$ denote the watermark decoder. For a FULL prompt carrying $\mathbf{m}$, we require
\begin{equation}
	D_s\!\left(D_{\phi}(F_{\theta}(p,\boldsymbol{\epsilon}),p)\right)
	\approx \mathbf{m},
	\qquad s(p)=\mathrm{FULL},
\end{equation}
where $\approx$ denotes recovery with a low bit error rate.

Second, the decoder should enforce the output quality associated with each access state while preserving the visual quality under full access. Let $D_{0}$ denote the original VAE decoder, and define $\mathbf{x}_{\mathrm{ref}}=D_{0}(\mathbf{z})$ as the reference decoding of the same latent representation. The intended decoding behavior is
\begin{equation}
	D_{\phi}(\mathbf{z},p)\approx
	\begin{cases}
		\mathbf{x}_{\mathrm{ref}},
		& s(p)=\mathrm{FULL},\\
		\mathcal{B}(\mathbf{x}_{\mathrm{ref}}),
		& s(p)=\mathrm{INCOMPLETE},\\
		\mathbf{x}_{\mathrm{black}},
		& s(p)=\mathrm{INVALID},
	\end{cases}
	\label{eq:decoder_behavior}
\end{equation}
where $\mathcal{B}$ is a blurring operator and $\mathbf{x}_{\mathrm{black}}$ is a uniformly black image. Here, $\approx$ denotes image fidelity to the corresponding reference output. Thus, full access preserves visual quality and watermark recoverability, preview access retains only coarse visual content, and no access suppresses image content. 

More generally, the preview behavior for trial users is specified through the supervision targets used during access-control training. The model owner can define these targets according to the service provider's requirements, aiming to retain limited visual quality for trial evaluation. Examples of alternative target transformations include monochrome conversion, color reduction, Sobel-based edge extraction, etc.


\begin{figure*}[!t]
	\centering
	\includegraphics[width=0.85\textwidth]{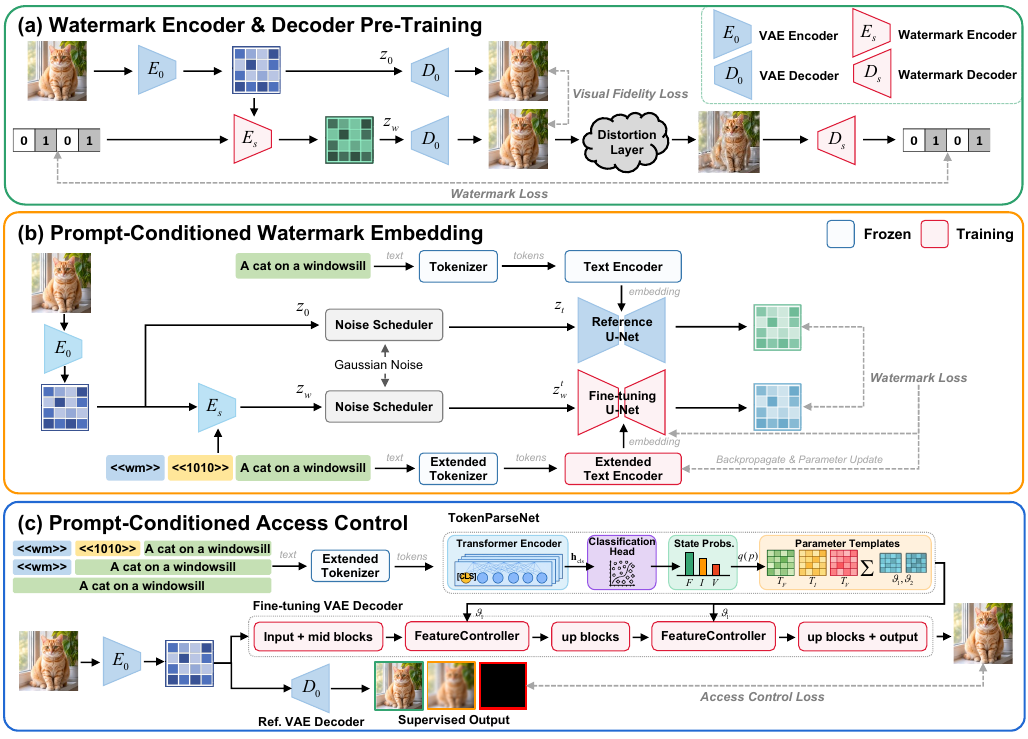}
	\caption{Overview of the proposed framework: (a) Watermark codec pretraining with visual fidelity and watermark recovery objectives; (b) prompt-conditioned watermark embedding through joint fine-tuning of the U-Net and protocol-token input embeddings; and (c) prompt-conditioned access control for differentiated output quality.}
	\label{fig:Methodology}
\end{figure*}

\subsection{Overall Pipeline}

To unify watermark embedding and access control through a shared prompt while facilitating stable optimization, we propose a two-stage training framework, outlined in Fig.~\ref{fig:Methodology}. The first stage enables the model to generate images carrying the watermark message specified by the prompt while preserving visual quality. The second stage regulates output quality according to the prompt's protocol structure, providing differentiated access levels. Together, the two stages allow a single prompt to specify watermark content and control access to high-quality generation.

In the \textbf{Prompt-Conditioned Watermark Embedding} stage, we first pretrain a watermark codec to provide robust and visually imperceptible reference watermark patterns for subsequent learning. We then extend the tokenizer vocabulary so that each trigger or watermark word is encoded as a single dedicated token rather than split into existing subword tokens, and initialize the corresponding input embeddings in the text encoder. Aiming to associate each prompt-encoded message with its corresponding watermark pattern in the generated images while preserving the model's original generation quality, we jointly optimize the U-Net's parameters and the newly introduced token embeddings.

In the \textbf{Prompt-Conditioned Access Control} stage, we introduce an auxiliary network that receives the token IDs produced by the extended tokenizer, recognizes the encoded protocol structure, and generates dynamic convolution parameters for trainable modules inserted into the VAE decoder. We train these modules together with the decoder's convolutional layers using supervision for the three prompt classes and their corresponding access states. This stage establishes differentiated access through decoder behavior while preserving the watermark-bearing content under full access.

\subsection{Prompt Protocol and Access States}
\label{sec:prompt_protocol}
To encode watermark messages and regulate access through a unified prompt, we define a protocol that represents the message $\mathbf{m}$ as an ordered sequence of watermark words and determines the access state $s(p)$ from the prompt structure. 

We first partition the 48-bit message into $x$ consecutive segments of $y$ bits each, such that $\mathbf{m}=\mathbf{b}_0\Vert\mathbf{b}_1\Vert\cdots\Vert\mathbf{b}_{x-1}$, where $\mathbf{b}_i\in\{0,1\}^{y}$ and $\Vert$ denotes bit concatenation. Each segment is represented by one designated watermark word. We denote this payload partitioning scheme by $x\mathrm{t}y\mathrm{b}$, with positive integers $x$ and $y$ satisfying $xy=48$.

Given a partitioning scheme, a shared mapping function $w$ converts each possible $y$-bit segment into a distinct textual watermark word. The resulting watermark vocabulary contains $2^y$ watermark words. We additionally introduce a designated trigger word $\tau=\texttt{<<wm>>}$ to identify the start of the watermark prefix in a FULL prompt. The extended tokenizer maps each of the $2^y$ watermark words and the trigger word to a distinct protocol token, yielding $2^y+1$ protocol tokens, each associated with a trainable embedding in the text encoder.

We adopt $12\mathrm{t}4\mathrm{b}$ as the default scheme to balance watermark performance, training efficiency, and image generation quality; alternative partitioning schemes are evaluated in the ablation study. Under this configuration, the message is divided into twelve 4-bit segments, and the protocol vocabulary comprises 16 watermark words and one trigger word. Each watermark word has the textual form \texttt{<<bbbb>>}, where \texttt{bbbb} specifies its four payload bits. For example, $w((1,0,1,0))=\texttt{<<1010>>}$.

The twelve watermark words are arranged in the same order as their corresponding segments, so concatenating their payloads recovers the complete message $\mathbf{m}$. The text encoder incorporates positional information into their representations, enabling the U-Net to condition watermark embedding on both token identity and segment position. The presence and arrangement of these protocol elements determine the prompt class and its associated access state, as defined below.

Let $W(\mathbf{m})=[w(\mathbf{b}_0),\ldots,w(\mathbf{b}_{11})]$ denote the ordered watermark-word sequence, and let $\oplus$ denote concatenation of prompt components with separating spaces. 
Together with the semantic caption $c$, we distinguish three prompt classes and their associated access states:

\textbf{1) FULL: full access.}
A FULL prompt follows the structure
\begin{equation}
	p_{\mathrm{full}}=\tau\oplus W(\mathbf{m})\oplus c.
\end{equation}

It contains exactly one trigger word at the beginning, immediately followed by twelve watermark words and then the semantic caption. Each watermark word may represent any 4-bit value, and concatenating their payloads in sequence yields the message $\mathbf{m}$. This state is intended to produce a clear, high-quality image consistent with $c$ and carrying the specified traceability watermark.

\textbf{2) INCOMPLETE: preview access.}
An INCOMPLETE prompt contains exactly one trigger word at the beginning, immediately followed by the semantic caption, with no watermark words and takes the form:
\begin{equation}
    p_{\mathrm{incomplete}}=\tau\oplus c.
\end{equation}

This prompt structure is assigned to the preview state for trial users, for which the decoder is trained to produce blurred images that retain coarse semantic content.

\textbf{3) INVALID: no access.}
An INVALID prompt contains neither the trigger nor any watermark word and therefore takes the form
\begin{equation}
	p_{\mathrm{invalid}}=c.
\end{equation}

It is assigned to the no-access state, for which the decoder is trained to produce a uniformly black image.

\subsection{Prompt-Conditioned Watermark Embedding}

This stage equips the generator with the ability to embed prompt-specified watermark messages during denoising. We first pretrain a watermark codec and extend the tokenizer and text embeddings, then transfer the learned watermark patterns into the U-Net through the paired reference and student branches illustrated in Fig.~\ref{fig:Methodology}(b). The following subsections describe these steps.

\subsubsection{Watermark Codec Pretraining}

In this training phase, we hope to train an image watermarking network that can generate stable watermark patterns and extract watermark information. Following the encode-decode network designs of AquaLoRA~\cite{feng2024aqualora} and AuthenLoRA~\cite{shi2025authenlora}, we jointly pretrain a latent-space watermark encoder $E_s$ and a pixel-space watermark decoder $D_s$, as illustrated in Fig.~\ref{fig:Methodology}(a). The pretrained codec provides a stable reference watermark pattern for subsequent prompt-conditioned fine-tuning of the U-Net and text encoder.

Given an image $\mathbf{x}\sim\mathcal{D}$ from the training dataset $\mathcal{D}$ and a randomly sampled watermark message $\mathbf{m}$, the frozen VAE encoder $E_0$ maps the image into a latent representation $\mathbf{z}$. The watermark encoder $E_s$ produces a message-dependent residual $\Delta\mathbf{z}$, which is added to $\mathbf{z}$ to obtain the watermarked latent $\mathbf{z}_{\mathrm{w}}$. The frozen VAE decoder $D_0$ then reconstructs the watermarked image:
\begin{equation}
	\begin{aligned}
		\mathbf{z} &= E_0(\mathbf{x}),
		& \Delta\mathbf{z} = E_s(\mathbf{z},\mathbf{m}),\\
		\mathbf{z}_{\mathrm{w}} &= \mathbf{z}+\Delta\mathbf{z},
		& \mathbf{x}_{\mathrm{w}} = D_0(\mathbf{z}_{\mathrm{w}}).
	\end{aligned}
\end{equation}
Here, $\Delta\mathbf{z}$ represents the latent-space watermark pattern learned by the encoder. A stochastic distortion layer $\mathcal{A}$ simulates common image-processing attacks, and the watermark decoder $D_s$ recovers the embedded message from the distorted image:
\begin{equation}
	\hat{\mathbf{m}}
	=
	D_s\!\left(\mathcal{A}(\mathbf{x}_{\mathrm{w}})\right).
\end{equation}

We optimize the codec for watermark decodability and visual fidelity. Binary cross-entropy between $\hat{\mathbf{m}}$ and $\mathbf{m}$ encourages accurate extraction under distortions. Visual fidelity is constrained using LPIPS~\cite{zhang2018lpips}, Peak Regional Variation Loss (PRVL)~\cite{feng2024aqualora} and Perceptual Residual Tone Suppression (PRTS) Loss~\cite{shi2025authenlora}, computed between $\mathbf{x}_{\mathrm{w}}$ and the reference reconstruction $\mathbf{x}_{\mathrm{ref}}=D_0(\mathbf{z})$. 

The overall codec pretraining objective is
\begin{equation}
	\begin{aligned}
		\mathcal{L}_{\mathrm{codec}}
		&= \mathcal{L}_{\mathrm{msg}}
		+ \lambda\mathcal{L}_{\mathrm{LPIPS}}+\mu\mathcal{L}_{\mathrm{PRVL}}
		+ \nu\mathcal{L}_{\mathrm{PRTS}},
	\end{aligned}
	\label{eq:codec_loss}
\end{equation}
where $\mathcal{L}_{\mathrm{msg}}$ is the expected binary cross-entropy over both watermarked samples with their embedded messages and unwatermarked samples with all-zero targets. The remaining terms denote the visual fidelity losses described above.

After pretraining, $E_s$ and $D_s$ are frozen to provide consistent watermark reference and extraction for subsequent stages.

\subsubsection{Tokenizer Vocabulary Extension}

To map textual watermark words to dedicated tokens with learnable input embeddings while preserving pretrained lexical representations, we extend the tokenizer vocabulary and the input embedding layer of the text encoder. Following Section~\ref{sec:prompt_protocol}, we construct the protocol vocabulary
\begin{equation}
	\mathcal{V}_{\mathrm{p}}
	=
	\{\tau\}
	\cup
	\{w(\mathbf{b}) \mid \mathbf{b}\in\{0,1\}^{4}\}.
\end{equation}

This vocabulary contains 16 watermark words and one trigger word. Each is assigned a unique token ID outside the original vocabulary and encoded as a single token, preventing its textual form from being fragmented into existing subword tokens.

Following this, we expand the text encoder's embedding matrix with a trainable block $\mathbf{U}\in\mathbb{R}^{K\times d}$, where $K=17$ and $d$ is the embedding dimension. During subsequent watermark fine-tuning, only the newly introduced token embeddings are updated within the text encoder; the original token embeddings and all remaining text encoder parameters are frozen. This preserves the pretrained lexical parameters while allowing the protocol tokens to acquire watermark-specific representations.

\subsubsection{Prompt-Conditioned Fine-Tuning}

We next transfer the pretrained watermark pattern into the denoising model and associate it with the protocol-token embeddings. The objective is to make the watermark message specified by a FULL prompt govern the watermark pattern in the generated latent representation. 

To achieve this, we align the student's noise prediction on a watermarked latent with the reference model's prediction on its clean counterpart. The reference branch provides the original denoising behavior under the semantic caption, while the student branch learns to accommodate the watermark displacement under the corresponding FULL prompt. This establishes a training signal for watermark-conditioned generation while encouraging preservation of the pretrained model's behavior.

Let $\epsilon_0$ denote the reference U-Net and $\epsilon_{\psi}$ the student U-Net, and let $T_0$ denote the frozen reference text encoder and $T_{\mathbf{U}}$ the extended text encoder whose only trainable parameters are the protocol-token embeddings $\mathbf{U}$. For an image-caption pair $(\mathbf{x},c)$ sampled from the training dataset and a random message $\mathbf{m}$, we construct
\begin{equation}
	\begin{aligned}
		\mathbf{z} &= E_0(\mathbf{x}),
		& \Delta\mathbf{z} &= E_s(\mathbf{z},\mathbf{m}),\\
		\mathbf{z}_{\mathrm{w}} &= \mathbf{z}+\Delta\mathbf{z},
		& p_{\mathrm{full}} &= \tau\oplus W(\mathbf{m})\oplus c.
	\end{aligned}
\end{equation}

We sample a timestep $t$ and Gaussian noise
$\boldsymbol{\epsilon}\sim\mathcal{N}(\mathbf{0},\mathbf{I})$,
and apply the same forward-noising process to both latent representations:
\begin{equation}
	\begin{aligned}
		\mathbf{z}_t
		&= \sqrt{\bar{\alpha}_t}\,\mathbf{z}
		+ \sqrt{1-\bar{\alpha}_t}\,\boldsymbol{\epsilon},\\
		\mathbf{z}^{\mathrm{w}}_t
		&= \sqrt{\bar{\alpha}_t}\,\mathbf{z}_{\mathrm{w}}
		+ \sqrt{1-\bar{\alpha}_t}\,\boldsymbol{\epsilon},
	\end{aligned}
\end{equation}
where $\bar{\alpha}_t$ is the cumulative noise-schedule coefficient. Sharing the noise realization ensures that the difference between the two noisy inputs arises solely from the watermark displacement. The reference and student predictions are
\begin{equation}
	\begin{aligned}
		\hat{\boldsymbol{\epsilon}}_{\mathrm{ref}}
		&= \epsilon_0(\mathbf{z}_t,t,T_0(c)),\\
		\hat{\boldsymbol{\epsilon}}_{\mathrm{w}}
		&= \epsilon_{\psi}
		(\mathbf{z}^{\mathrm{w}}_t,t,T_{\mathbf{U}}(p_{\mathrm{full}})).
	\end{aligned}
\end{equation}
We optimize their mean-squared discrepancy:
\begin{equation}
	\mathcal{L}_{\mathrm{watermark}}
	=
	\mathbb{E}_{(\mathbf{x},c),\mathbf{m},t,\boldsymbol{\epsilon}}
	\left[
	\frac{1}{N_z}
	\left\|
	\hat{\boldsymbol{\epsilon}}_{\mathrm{w}}
	-
	\operatorname{sg}
	(\hat{\boldsymbol{\epsilon}}_{\mathrm{ref}})
	\right\|_2^2
	\right],
\end{equation}
where $N_z$ is the number of latent elements. The stop-gradient operator $\operatorname{sg}$ treats the reference prediction as a fixed supervision target, preventing gradients from propagating through the reference branch.

This objective encourages the student to retain the watermark as part of the underlying signal rather than remove it as noise. Specifically, let
$\Delta\mathbf{z}=\mathbf{z}_{\mathrm{w}}-\mathbf{z}$
denote the pretrained watermark displacement. The paired noisy inputs satisfy
\begin{equation}
	\mathbf{z}^{\mathrm{w}}_t-\mathbf{z}_t
	=
	\sqrt{\bar{\alpha}_t}\,\Delta\mathbf{z}.
\end{equation}
For an $\epsilon$-prediction model, the estimated clean latent is obtained by subtracting the predicted noise contribution from the noisy input:
\begin{equation}
	\begin{aligned}
		\hat{\mathbf{z}}_{\mathrm{ref}}
		&=
		\frac{\mathbf{z}_t-
			\sqrt{1-\bar{\alpha}_t}\,
			\hat{\boldsymbol{\epsilon}}_{\mathrm{ref}}}
		{\sqrt{\bar{\alpha}_t}},\\
		\hat{\mathbf{z}}_{\mathrm{w}}
		&= \frac{\mathbf{z}^{\mathrm{w}}_t-
			\sqrt{1-\bar{\alpha}_t}\,
			\hat{\boldsymbol{\epsilon}}_{\mathrm{w}}}
		{\sqrt{\bar{\alpha}_t}}.
	\end{aligned}
\end{equation}
Consequently,
\begin{equation}
	\hat{\mathbf{z}}_{\mathrm{w}}-\hat{\mathbf{z}}_{\mathrm{ref}}
	=
	\Delta\mathbf{z}
	-
	\sqrt{\frac{1-\bar{\alpha}_t}{\bar{\alpha}_t}}
	\left(
	\hat{\boldsymbol{\epsilon}}_{\mathrm{w}}
	-
	\hat{\boldsymbol{\epsilon}}_{\mathrm{ref}}
	\right).
\end{equation}

At a fixed timestep, reducing the prediction discrepancy therefore makes the difference between the reconstructed latent estimates approach $\Delta\mathbf{z}$. Gradients jointly update the U-Net's parameters $\psi$ and the protocol-token embeddings $\mathbf{U}$, while the reference models, VAE, and watermark codec remain frozen.

After fine-tuning, a FULL prompt guides the adapted model to generate latent representations carrying the specified watermark pattern. Decoding these representations with the original VAE decoder $D_0$ yields images from which the watermark decoder $D_s$ can recover the message $\mathbf{m}$, while maintaining semantic consistency with the caption $c$ and visual quality comparable to that of the reference model. 

\subsection{Prompt-Conditioned Access Control}

In this stage, we aim to equip the VAE decoder with the ability to regulate output quality according to the prompt protocol while preserving watermark-bearing content under full access. As illustrated in Fig.~\ref{fig:Methodology}(c), TokenParseNet interprets the prompt structure and generates control parameters for two FeatureController modules within the decoder. We first describe these components, then present the training procedure that associates the three access states with their intended decoding behavior.

\subsubsection{TokenParseNet and FeatureController}
TokenParseNet maps the tokenized prompt to dynamic convolution parameters through learned state prediction. It comprises a Transformer encoder with a state classification head and a trainable parameter template $\mathbf{T}$. The encoder captures the prompt structure, while the template translates the predicted state probabilities into modulation parameters.

TokenParseNet receives the token IDs produced by the extended tokenizer together with the corresponding attention mask. Its own trainable embedding layer, independent of the text encoder, maps these IDs to token embeddings. A learnable classification-token embedding, denoted by \texttt{cls}, is prepended to the embedding sequence. The classification token aggregates information across the sequence through self-attention, providing a dedicated representation for prompt-level state prediction. This enables the classifier to consider the presence and arrangement of protocol tokens jointly.

A two-layer Transformer encoder processes the embedding sequence. The resulting contextualized classification-token representation $\mathbf{h}_{\mathrm{cls}}$ is used to obtain the state probabilities:
\begin{equation}
	\mathbf{q}(p)
	=
	\operatorname{softmax}
	\left(
	\mathbf{W}_{\mathrm{cls}}
	\operatorname{LN}(\mathbf{h}_{\mathrm{cls}})
	+
	\mathbf{b}_{\mathrm{cls}}
	\right).
\end{equation}
Here, $\operatorname{LN}$ denotes layer normalization, and
$\mathbf{W}_{\mathrm{cls}}\in\mathbb{R}^{3\times d}$ and
$\mathbf{b}_{\mathrm{cls}}\in\mathbb{R}^{3}$ are the learnable
weight matrix and bias vector of the state classification head,
respectively, where $d$ is the dimension of
$\mathbf{h}_{\mathrm{cls}}$. Each component $q_s(p)$ corresponds to a state
$s\in\mathcal{S}$, where
$\mathcal{S}=\{\mathrm{FULL},\mathrm{INCOMPLETE},\mathrm{INVALID}\}$.

For each modulation site $\ell\in\{1,2\}$, the template contains a trainable parameter vector $\mathbf{T}_{s,\ell}$ for each state. The predicted probabilities combine these templates to produce the corresponding control parameters:
\begin{equation}
	\boldsymbol{\vartheta}_{\ell}(p)
	=
	\sum_{s\in\mathcal{S}}
	q_s(p)\mathbf{T}_{s,\ell}.
\end{equation}

Each parameter vector is supplied to its corresponding FeatureController, which applies prompt-dependent convolution to the intermediate features and adds the resulting residual to the original features. For $512\times512$ outputs, the two modules operate at spatial resolutions of $64\times64$ and $256\times256$, respectively. The first follows the original decoder's middle block, and the second follows its second upsampling block. Together, this differentiable modulation connects prompt structure to decoding behavior, while retaining the original decoding architecture. 

\subsubsection{Training Procedure}
We initialize the controlled decoder from the original VAE decoder, ensuring a stable start for training. We jointly train TokenParseNet, both FeatureController modules, and all convolutional layers within the decoder's middle and upsampling blocks. All remaining parameters of the original decoder are frozen, as are the reference VAE and pretrained watermark codec.

At each training step, we independently sample an access state $s$ for each image-caption pair $(\mathbf{x},c)$ according to predefined probabilities for FULL, INCOMPLETE, and INVALID. We then sample a watermark message $\mathbf{m}$ and construct the corresponding prompt $p$ following Section~\ref{sec:prompt_protocol}. The state sampling probabilities and their schedule are specified in the Experimental Setup section.

Training latents are obtained directly from the frozen VAE encoder and watermark encoder:
\begin{equation}
	\begin{aligned}
		\mathbf{z} &= E_0(\mathbf{x}),\\
		\widetilde{\mathbf{z}}
		&=
		\begin{cases}
			\mathbf{z}+E_s(\mathbf{z},\mathbf{m}),
			& s=\mathrm{FULL},\\
			\mathbf{z},
			& s\in\{\mathrm{INCOMPLETE},\mathrm{INVALID}\}.
		\end{cases}
	\end{aligned}
\end{equation}
The reference decoder and controlled decoder receive the same latent $\widetilde{\mathbf{z}}$, producing
\begin{equation}
	\mathbf{x}_{\mathrm{ref}}=D_0(\widetilde{\mathbf{z}}),
	\qquad
	\widehat{\mathbf{x}}=D_{\phi}(\widetilde{\mathbf{z}},p),
\end{equation}
where $\widehat{\mathbf{x}}$ denotes the output image of the controlled decoder. 

Following Eq.~\eqref{eq:decoder_behavior}, we define the supervision target $\mathbf{y}_s$ according to the access state: FULL uses the clear, watermark-bearing reference reconstruction $\mathbf{x}_{\mathrm{ref}}$; INCOMPLETE uses its blurred version; and INVALID uses a uniformly black image. These targets explicitly supervise the decoder to preserve image quality under full access, provide blurred previews under incomplete access, and suppress image content under invalid access.

The image objective combines pixel-wise fidelity across all states with perceptual fidelity for FULL samples:
\begin{equation}
	\begin{aligned}
		\mathcal{L}_{\mathrm{img}}
		=
		\mathbb{E}\Big[
		&\frac{1}{N_x}
		\left\|\widehat{\mathbf{x}}-\mathbf{y}_{s}\right\|_1
		+
		\mathbf{1}_{\{s=\mathrm{FULL}\}}
		\mathcal{L}_{\mathrm{LPIPS}}
		\left(\widehat{\mathbf{x}},\mathbf{y}_{s}\right)
		\Big],
	\end{aligned}
\end{equation}
where $N_x$ is the number of image elements and $\mathbf{1}_{\{\cdot\}}$ is the indicator function. Both terms are averaged over the entire image batch, so the sampled state proportions determine their relative contributions.

We additionally supervise TokenParseNet using the known state labels:
\begin{equation}
	\mathcal{L}_{\mathrm{state}}
	=
	-\frac{1}{|\mathcal{P}|}
	\sum_{p\in\mathcal{P}}
	\log q_{s(p)}(p),
\end{equation}
where $\mathcal{P}$ comprises the prompts from the image batch and an auxiliary batch containing approximately equal numbers of prompts from the three states. These auxiliary prompts require only TokenParseNet evaluation, providing additional structural supervision without image decoding. The overall objective is
\begin{equation}
	\mathcal{L}_{\mathrm{control}}
	=
	\mathcal{L}_{\mathrm{img}}
	+
	\lambda_{\mathrm{state}}\mathcal{L}_{\mathrm{state}}.
\end{equation}

At inference, TokenParseNet generates control parameters by weighting the learned templates according to the predicted state probabilities. The FeatureController modules use these parameters to adjust decoder features and regulate the quality of the output image.

\section{Experiments}
\subsection{Experimental Setup}
\subsubsection{Implementation Details}

For prompt-conditioned watermark embedding, we use Stable Diffusion v1.5 and jointly optimize U-Net LoRA adapters with rank 320 and scaling factor 1.0, together with the 17 newly introduced token embeddings in the text encoder. We use 10,000 samples from the Gustavosta/Stable-Diffusion-Prompts dataset. Training uses a batch size of 4 for 160 epochs. We employ AdamW with an initial learning rate of $10^{-4}$ for both parameter groups.

For prompt-conditioned access control, we jointly optimize TokenParseNet, both FeatureController modules, and the convolutional layers within the VAE decoder's middle and upsampling blocks. We train for 15,000 steps with a batch size of 2, using AdamW as the optimizer. Learning rates are set to $10^{-4}$ for TokenParseNet and $10^{-5}$ for the trainable decoder parameters. The FULL/INCOMPLETE/INVALID sampling probabilities are $(0.50,0.25,0.25)$ for the first 2,000 steps and $(0.25,0.50,0.25)$ thereafter. Each step additionally uses 32 auxiliary prompts for state classification, and $\lambda_{\mathrm{state}}$ is set to 1. All training and evaluation were conducted on a single NVIDIA GeForce RTX 4090 GPU with 24 GB of VRAM.
\subsubsection{Evaluation Metrics}
We evaluate the proposed method in terms of visual fidelity, text-image alignment, and watermark performance.

\textbf{Visual fidelity and text-image alignment.}
We use Fr\'echet Inception Distance (FID)~\cite{heusel2017fid}, CLIP similarity~\cite{radford2021clip}, and Learned Perceptual Image Patch Similarity (LPIPS)~\cite{zhang2018lpips} to assess the impact of watermark embedding. FID measures the distributional discrepancy between watermarked images and images generated by the original model. LPIPS measures perceptual differences between image pairs generated using the same semantic caption and initial noise. Lower FID and LPIPS indicate smaller deviations from the reference outputs. CLIP similarity measures the cosine similarity between the CLIP embeddings of each generated image and its semantic caption $c$, excluding protocol tokens; higher values indicate better text-image alignment.

To evaluate access control, we additionally report Peak Signal-to-Noise Ratio (PSNR)~\cite{wang2009mse} and Structural Similarity (SSIM)~\cite{wang2004ssim} between the controlled and reference VAE outputs decoded from the same latent representation. We report these metrics separately for FULL, INCOMPLETE, and INVALID prompts. High similarity under FULL prompts indicates preservation of decoding fidelity, whereas reduced similarity under INCOMPLETE and INVALID prompts characterizes the intended departure from unrestricted decoding.

\textbf{Watermark performance.}
We report bit accuracy (BitAcc) and true positive rate (TPR). BitAcc is the proportion of correctly recovered bits across all evaluated watermarked images, obtained by comparing the binary messages predicted by the watermark decoder $D_s$ with the embedded messages.

TPR is the proportion of watermarked images whose number of correctly recovered bits meets a detection threshold $\kappa$. We select this threshold at a target false positive rate (FPR) of $10^{-6}$. Under the null hypothesis of independent, uniformly random bit predictions, the number of matching bits follows $C\sim\mathrm{Binomial}(L,0.5)$, giving
\begin{equation}
	\kappa=\min\left\{k:\Pr(C\geq k)\leq 10^{-6}\right\}.
\end{equation}
For our 48-bit payload, $\kappa=41$; an image is therefore detected as watermarked if at least 41 bits match the specified message. Thus, BitAcc measures message recovery accuracy, while TPR measures the proportion of watermarked images that satisfy the detection criterion.

\begin{table*}[!t]
	\centering
	\caption{Comparison of watermark capacity, visual fidelity, text-image alignment, and watermark performance. Adv.\ denotes evaluation after common image-processing attacks.}
	\label{tab:watermark_comparison}
	\begin{tabular}{lcccccccc}
		\hline
		\multirow{2}{*}{\textbf{Methods}}
		& \multirow{2}{*}{\textbf{Capacity}}
		& \multicolumn{3}{c}{\textbf{Fidelity}}
		& \multirow{2}{*}{{BitAcc} $\uparrow$}
		& \multirow{2}{*}{{TPR} $\uparrow$}
		& \multirow{2}{*}{{BitAcc (Adv.)} $\uparrow$}
		& \multirow{2}{*}{{TPR (Adv.)} $\uparrow$} \\
		\cline{3-5}
		& & FID $\downarrow$
		& CLIP $\uparrow$
		& LPIPS $\downarrow$
		& & & & \\
		\hline
		\textit{Post-Processing Watermarking Methods} \\
		DWT-DCT
		& 48 bits & 3.4568 & 0.3183 & 0.0162
		& 0.8432 & 0.6500 & 0.5472 & 0.1038 \\
		DWT-DCT-SVD
		& 48 bits & 9.3839 & 0.3166 & 0.0237
		& 1.0000 & 1.0000 & 0.6650 & 0.2563 \\
		RivaGAN
		& 32 bits & 13.9262 & 0.3208 & 0.0467
		& 0.9899 & 0.9675 & 0.8672 & 0.6266 \\
		\hline
		\textit{During-Generation Watermarking Methods}\\
		Tree-Ring
		& 1 bit & 83.4464 & 0.3246 & 0.4402
		& -- & 1.0000 & -- & 0.9011 \\
		Stable Signature
		& 48 bits & 16.4728 & 0.3244 & 0.0492
		& 0.9943 & 1.0000 & 0.7955 & 0.5200 \\
		AquaLoRA
		& 48 bits & 69.8794 & 0.3226 & 0.2827
		& 0.9584 & 0.9525 & 0.8975 & 0.7711 \\
		\textbf{Ours (FULL Prompt)}
		& 48 bits & 71.2718 & 0.3235 & 0.3191
		& 0.9790 & 0.9900 & 0.9010 & 0.8227 \\
		\hline
	\end{tabular}
\end{table*}

\subsection{Watermark Performance and Visual Fidelity}
We compare the proposed method with representative watermarking baselines to assess the trade-off between watermark reliability and image fidelity. The evaluation considers payload capacity, visual fidelity, text-image alignment, and watermark recovery and detection under both unmodified and attacked conditions.

The baselines comprise two categories: post-processing watermarking methods, including DWT-DCT, DWT-DCT-SVD, and RivaGAN~\cite{zhang2019rivagan}, and during-generation watermarking methods, including Tree-Ring~\cite{wen2023treering}, Stable Signature~\cite{fernandez2023stable}, and AquaLoRA~\cite{feng2024aqualora}. We evaluate our method using the complete generation pipeline: FULL prompts guide the watermark-adapted U-Net to generate watermarked latent representations, which are then decoded into pixel-space images by the fine-tuned, access-controlled VAE decoder. This evaluation accounts for any potential degradation in image quality introduced by the adapted decoder.

\begin{figure}[!t]
	\centering
	\includegraphics[
	trim={12.79mm 377.9mm 0mm 0mm},clip,width=\linewidth]{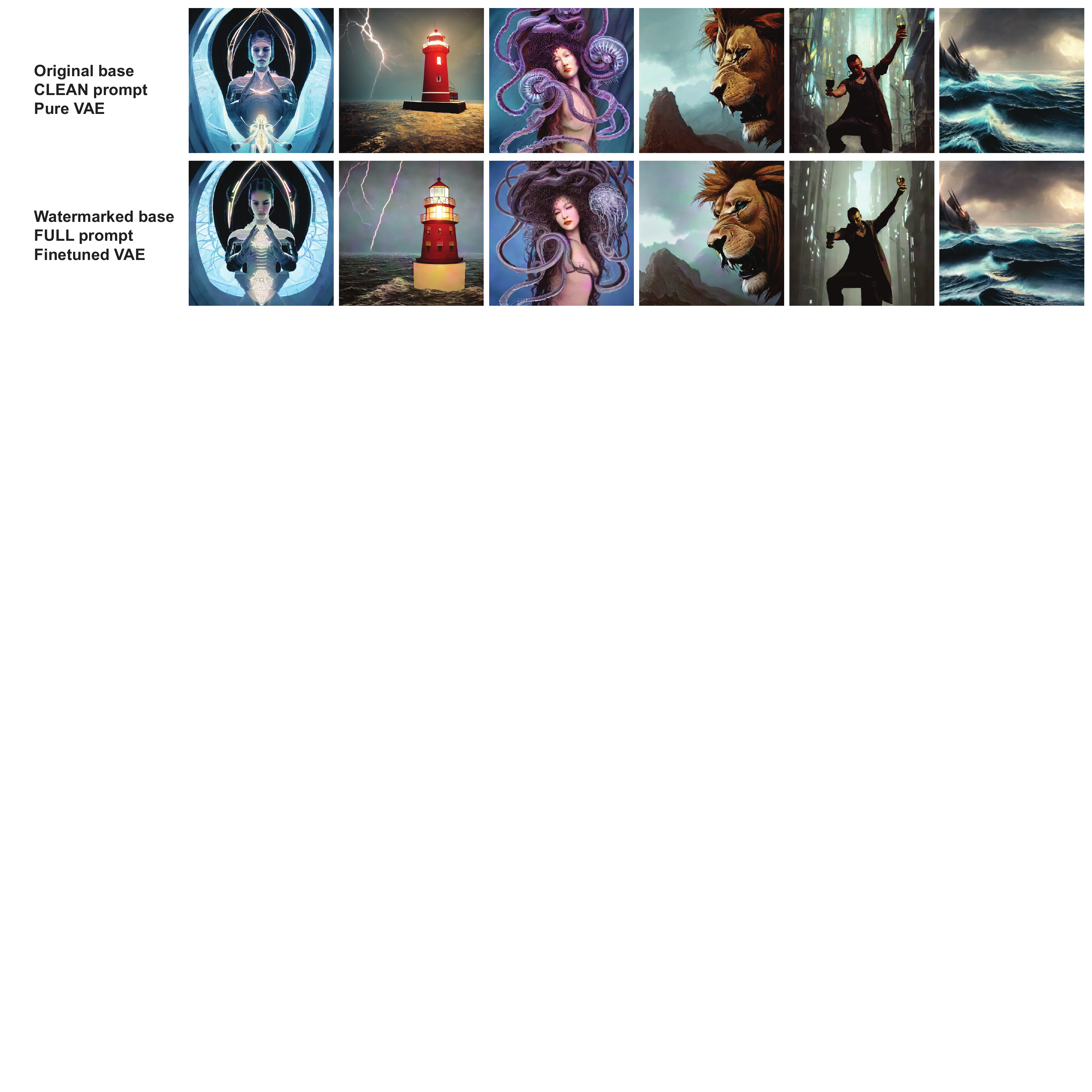}
	\caption{Qualitative comparison of reference and watermarked outputs. Top: the original base model and VAE decoder with clean prompts; bottom: the watermark-adapted base model and access-controlled VAE decoder with corresponding FULL prompts.}
	\label{fig:visual_fidelity}
\end{figure}

Visual fidelity is measured against the corresponding reference images, while text-image alignment and watermark performance are evaluated on the watermarked outputs. For robustness evaluation, denoted by ``Adv.'', we apply common image-processing operations, including color jitter, Gaussian blurring, JPEG compression, etc., and measure BitAcc and TPR on the resulting images.

Table~\ref{tab:watermark_comparison} compares the seven methods across these dimensions. Post-processing methods achieve the best visual fidelity, with DWT-DCT-SVD additionally attaining perfect BitAcc and TPR without attacks. However, their detection performance deteriorates under the evaluated image-processing operations, indicating limitations when images undergo potential attacks. Our method achieves a BitAcc of 0.9790 and a TPR of 0.9900 without attacks, and the highest attacked BitAcc of 0.9010 and TPR of 0.8227 among the evaluated multi-bit methods. Compared with AquaLoRA at the same payload capacity, our method achieves higher BitAcc and TPR under both attack-free and attacked conditions, at the cost of a slight reduction in visual fidelity. The FID and LPIPS of our proposed method are lower than those of Tree-Ring, although Tree-Ring retains a higher attacked TPR in its detection-only setting. Overall, our proposed method offers a favorable balance between 48-bit watermark performance, robustness, and text-image alignment.

Fig.~\ref{fig:visual_fidelity} provides a qualitative comparison between the original generation pipeline and our complete framework. The first row shows reference images generated by the original base model and VAE decoder using clean prompts. The second row shows images generated by the watermark-adapted base model and access-controlled VAE decoder using FULL prompts with the corresponding semantic captions. Our outputs retain the main subjects, overall composition, and visual style of the reference images. These examples suggest that our proposed framework supports watermark embedding and access control while preserving the model's ability to generate visually coherent and detailed images under full access.

\begin{figure*}[!t]
	\centering
	\includegraphics[trim={0mm 137.47mm 2mm 0mm},clip,width=1\linewidth]{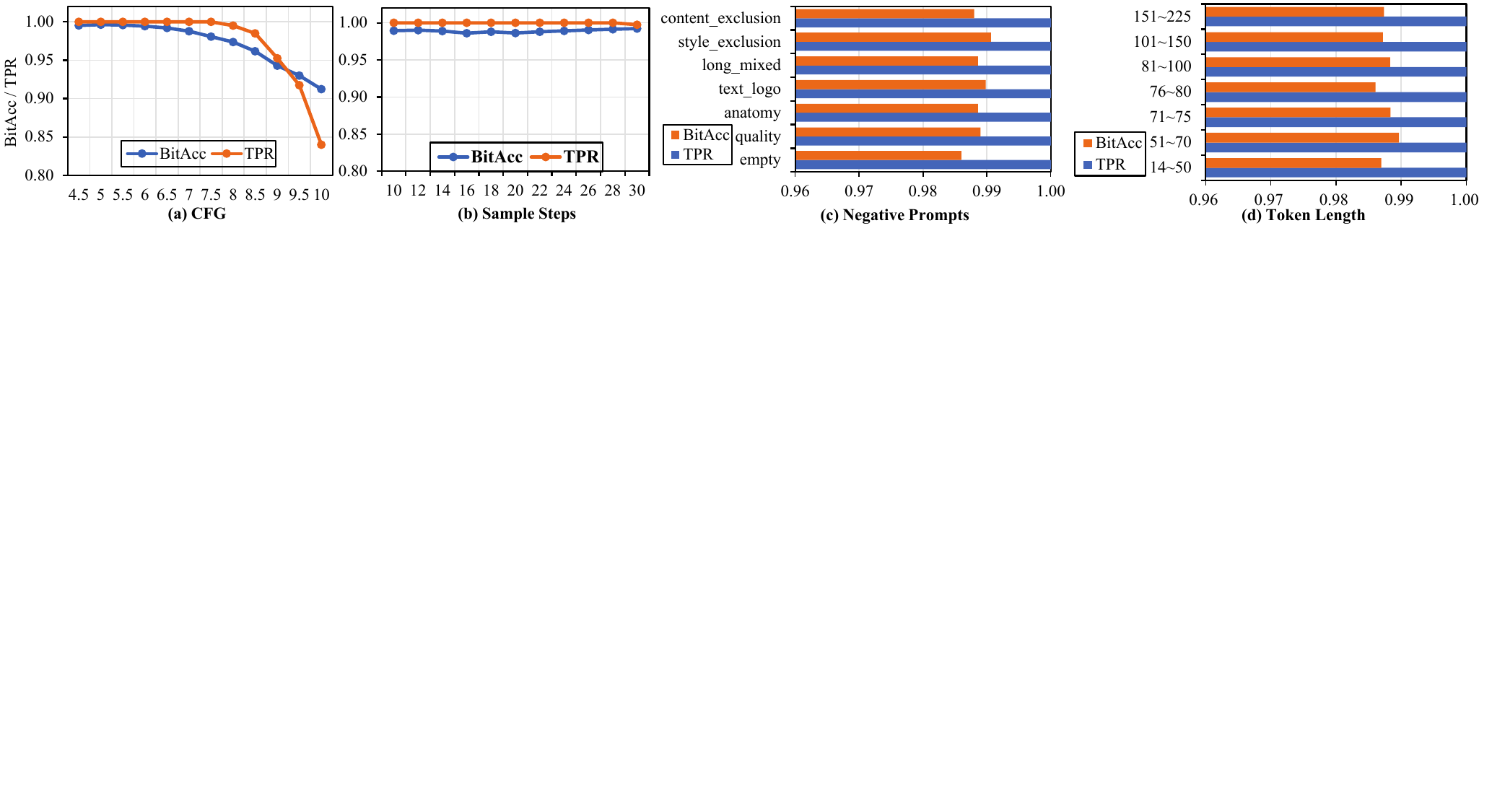}
	\caption{Watermark performance under different generation settings: (a) classifier-free guidance (CFG) scale, (b) number of sampling steps, (c) negative prompt category, and (d) sequence length of tokenized FULL prompts.}
	\label{fig:universality}
\end{figure*}

\subsection{Performance of Access Control}

In this experiment, we evaluate whether the fine-tuned VAE decoder preserves high-quality generation under FULL prompts while producing the intended blurred previews and black images under INCOMPLETE and INVALID prompts, respectively.

\begin{table}[!t]
	\centering
	\caption{Visual fidelity of PCDiff and our method under respective evaluation conditions. -- denotes an unreported metric.}
	\label{tab:prompt_comparison}
	\begin{tabular}{@{}llcccc@{}}
		\hline
		\multirow{2}{*}{\textbf{Method}}
		& \multirow{2}{*}{\textbf{Condition}}
		& \multicolumn{4}{c}{\textbf{Visual Fidelity}} \\
		\cline{3-6}
		& & PSNR & SSIM & LPIPS & FID \\
		\hline
		
		\multirow{3}{*}{\shortstack{PCDiff\\(86)}}
		& ori
		& 28.0088 & 0.9076 & -- & 8.5121 \\
		& wrong key
		& 24.6234 & 0.8664 & -- & 36.2917 \\
		& no fuser
		& 22.1754 & 0.8340 & -- & 36.6722 \\
		\hline
		
		\multirow{3}{*}{\shortstack{PCDiff\\(204)}}
		& ori
		& 30.1266 & 0.9095 & -- & 8.5375 \\
		& wrong key
		& 25.3485 & 0.8533 & -- & 34.8126 \\
		& no fuser
		& 21.4896 & 0.8221 & -- & 39.3671 \\
		\hline
		
		\multirow{3}{*}{\shortstack{PCDiff\\(208)}}
		& ori
		& 28.4268 & 0.8778 & -- & 11.2761 \\
		& wrong key
		& 22.2775 & 0.8134 & -- & 44.2957 \\
		& no fuser
		& 17.2655 & 0.7316 & -- & 63.8445 \\
		\hline
		
		\multirow{3}{*}{\textbf{Ours}}
		& FULL
		& 45.0988 & 0.9944 & 0.0016 & 0.8139 \\
		& INCOMPLETE
		& 21.2710 & 0.5973 & 0.6852 & 174.9066 \\
		& INVALID
		& 6.4560 & 0.0104 & 1.0084 & 448.3731 \\
		\hline
	\end{tabular}
\end{table}

Table~\ref{tab:prompt_comparison} presents the visual fidelity results of our method alongside those reported for PCDiff~\cite{gai2025pcdiff}. For our method, fidelity is evaluated by comparing the outputs of the controlled and original VAE decoders given the same latent representations for each prompt class. Under FULL prompts, high PSNR and SSIM, together with low LPIPS and FID, indicate that the controlled decoder closely preserves the visual fidelity of the original decoder. Under INCOMPLETE and INVALID prompts, PSNR and SSIM decrease substantially, while LPIPS and FID increase, reflecting pronounced deviations from the reference outputs, consistent with the intended restrictions on generation quality.

\begin{figure}
	\centering
	\includegraphics[width=\linewidth]{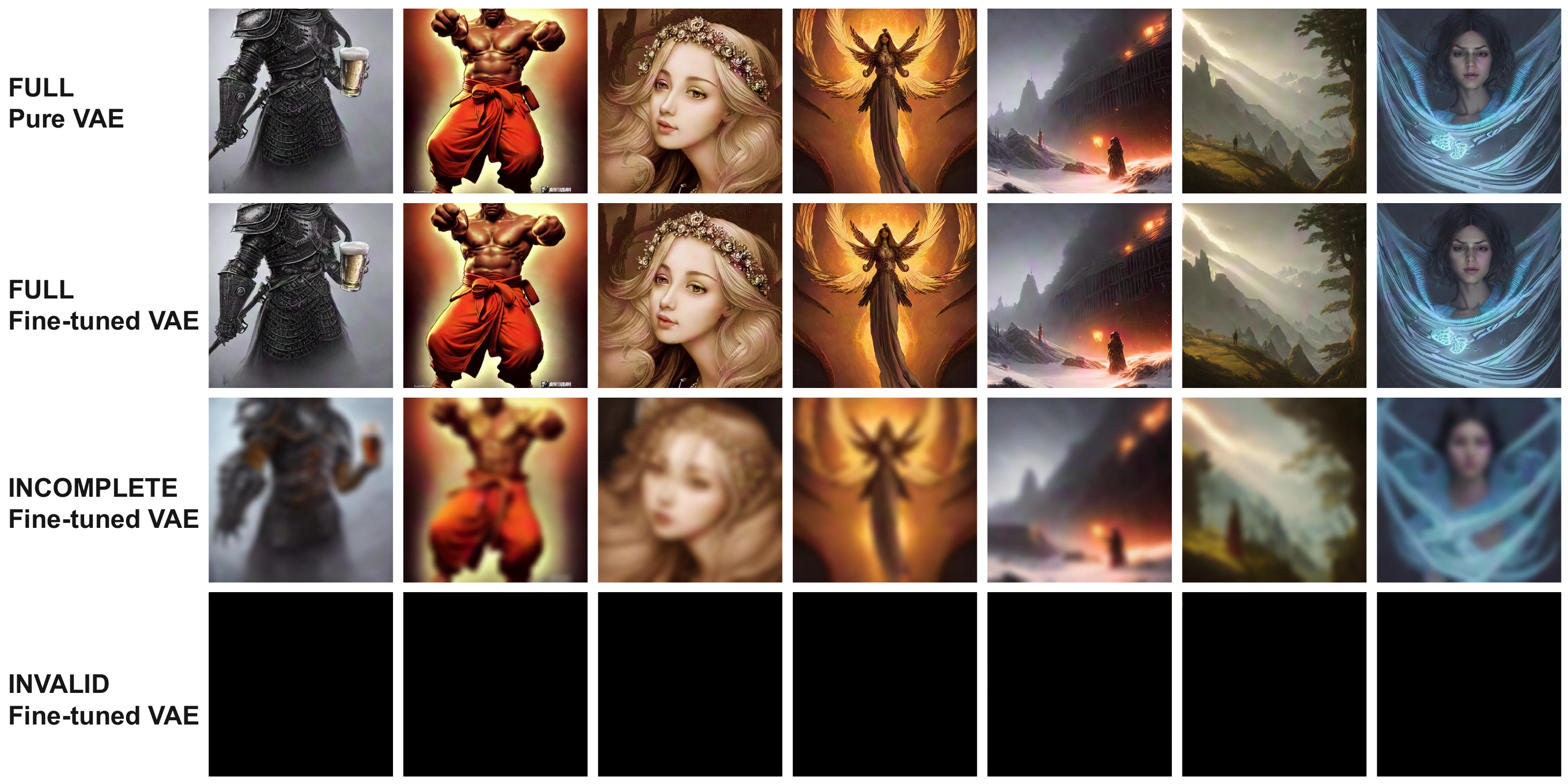}
	\captionof{figure}{Qualitative results of prompt-conditioned access control under different prompt classes.}
	\label{fig:access_control_visual}
\end{figure}

Compared with the reported results of PCDiff~\cite{gai2025pcdiff}, Table~\ref{tab:prompt_comparison} shows a more pronounced separation in output fidelity across the three conditions of our method. This separation is encouraged by explicitly supervising the decoder to produce distinct outputs for different prompt classes during training, thereby preserving clear images under full access while enforcing blurred previews or black outputs under restricted access. These results supports service tiers with clearly differentiated access to generated content.

Fig.~\ref{fig:access_control_visual} further illustrates these distinct decoding behaviors. The first two rows compare the original and fine-tuned VAE decoders under FULL prompts, showing nearly identical visual content and fine details. The third row presents blurred outputs for INCOMPLETE, retaining coarse semantic content while suppressing fine detail. The final row shows black outputs under INVALID prompts, effectively suppressing visible image content. Together, the quantitative and qualitative results support the effectiveness of prompt-conditioned access control across the evaluated cases while demonstrating that decoding fidelity is preserved under full access.

\subsection{Universality}
Generation settings and parameters can influence watermark performance. To assess the generalizability of the proposed watermarking method across inference configurations, we evaluate its performance under different classifier-free guidance (CFG) scales, sampling steps, as well as different types of negative prompts and the length of FULL prompts measured after tokenization.

As shown in Fig.~\ref{fig:universality}(a), BitAcc remains higher than 90\% as the CFG scale varies from 4.5 to 10, indicating that watermark remains effective across a wide range of CFG. Additionally, Fig.~\ref{fig:universality}(b) shows stable performance across 10--30 sampling steps, with both BitAcc and TPR exceeding 98\% throughout. These results demonstrate consistent watermark performance under the evaluated different generation parameters.

We further investigate the effects of negative prompt categories and FULL prompt length on watermark performance. We consider seven negative prompt categories: empty, quality, anatomy, text/logo, long mixed, style exclusion, and content exclusion. As shown in Fig.~\ref{fig:universality}(c), BitAcc remains above 98\% and TPR stays close to 100\% across all categories. To assess whether longer semantic descriptions interfere with watermark embedding, we retain the complete watermark-word prefix and vary the length of the subsequent semantic caption. Fig.~\ref{fig:universality}(d) shows that BitAcc consistently exceeds 98\% and TPR remains close to 100\% as the length of tokenized FULL prompts varies from 14 to 225 tokens. These results demonstrate consistent watermark performance across diverse prompt lengths and negative prompt configurations.

\subsection{Robustness of Watermarked Model}

In practical deployment, model providers may compress or adapt a watermarked model to meet resource constraints or customization requirements, which may unintentionally disrupt watermark generation. We therefore evaluate whether the proposed method continues to generate images with reliable watermarks after model modifications, including pruning, quantization and merging with randomly initialized or publicly available stylized LoRA adapters.

\begin{figure}[!t]
	\centering
	\includegraphics[trim={0mm 68.51mm 209.04mm 0mm},clip,width=0.9\linewidth]{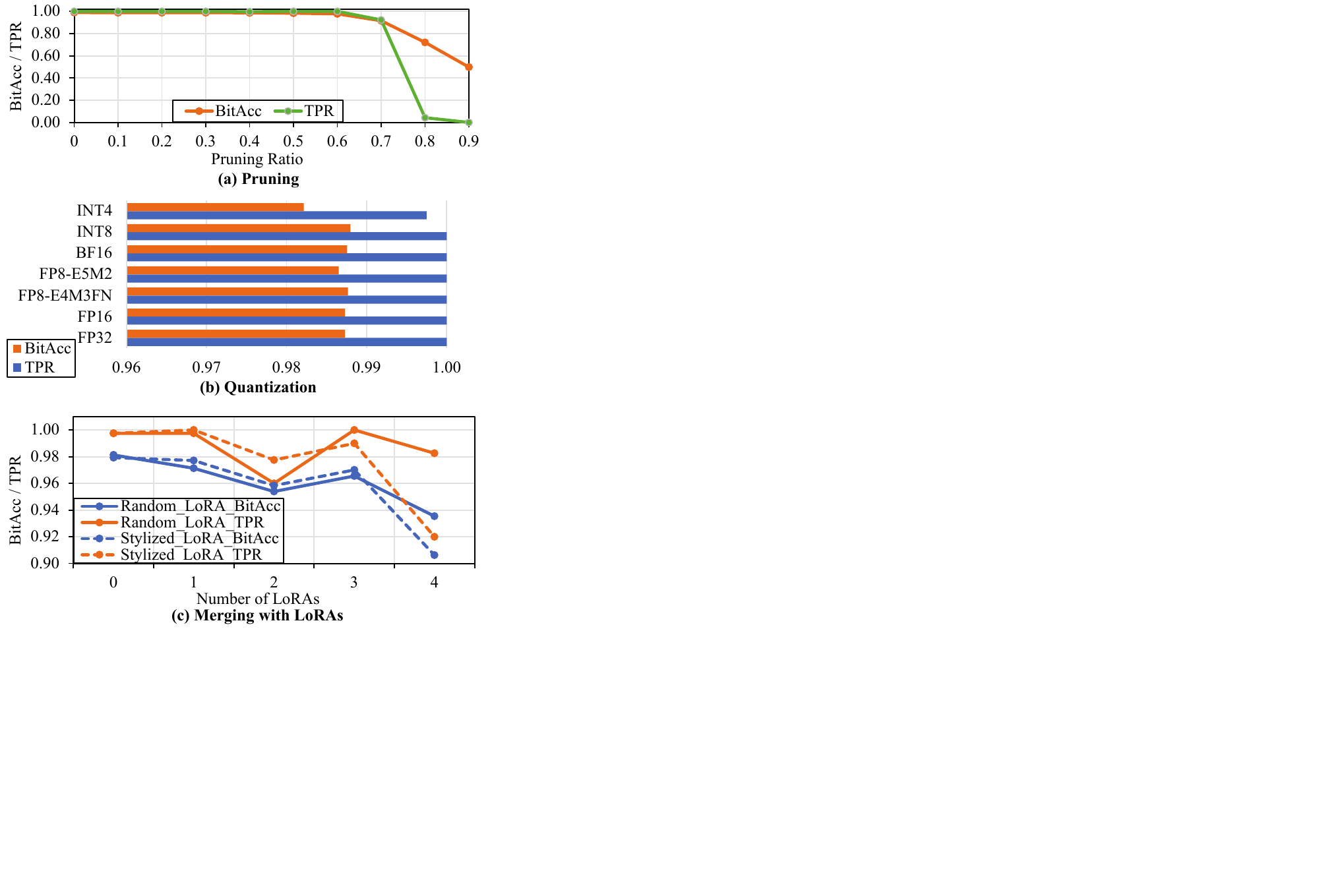}
	\caption{Watermark robustness under model modifications: (a) pruning, (b) quantization and (c) merging with random or style-specific LoRA models.}
	\label{fig:model_robustness}
\end{figure}

\begin{table}[!t]
	\centering
	\caption{Quantitative evaluation of image restoration attacks on blurred outputs generated under INCOMPLETE prompts.}
	\label{tab:restoration_attacks}
	\resizebox{\columnwidth}{!}{%
		\begin{tabular}{lcccc}
			\hline
			\multirow{2}{*}{\textbf{Method}}
			& \multicolumn{4}{c}{\textbf{Visual Fidelity}} \\
			\cline{2-5}
			& PSNR $\uparrow$ & SSIM $\uparrow$
			& FID $\downarrow$ & LPIPS $\downarrow$ \\
			\hline
			Blurred
			& 21.5183 & 0.6309 & 104.8734 & 0.6663 \\
			\hline
			Real-ESRGAN~\cite{wang2021realesrgan}
			& 21.4681 & 0.6272 & 110.2297 & 0.6589 \\
			SwinIR~\cite{liang2021swinir}
			& 21.8981 & 0.6370 & 86.0989 & 0.5748 \\
			Restormer~\cite{zamir2022restormer}
			& 21.4676 & 0.6280 & 109.3306 & 0.6771 \\
			DiffBIR~\cite{lin2024diffbir}
			& 22.4619 & 0.6420 & 76.0843 & 0.5535 \\
			\hline
		\end{tabular}%
	}
\end{table}

\subsubsection{Pruning}
We apply layer-wise magnitude pruning with pruning ratios ranging from 0 to 0.9. Within each layer, weights are ranked by their absolute values, and the specified proportion of the smallest-magnitude weights is set to zero. As shown in Fig.~\ref{fig:model_robustness}(a), both BitAcc and TPR remain near perfect at pruning ratios up to 0.7, demonstrating resilience to substantial parameter sparsification. 

\subsubsection{Quantization}
We evaluate watermark performance across seven numerical formats: FP32, FP16, BF16, FP8-E4M3FN, FP8-E5M2, INT8, and INT4. As illustrated in Fig.~\ref{fig:model_robustness}(b), the proposed method maintains high BitAcc and near-perfect TPR across all evaluated formats. Even under low-precision INT4 quantization, BitAcc remains above 98\% and TPR stays close to 100\%, confirming that the watermark remains fully detectable even under aggressive compression.

\subsubsection{Merging LoRAs}
To simulate provider-side model customization for personalized image generation, we separately merge the watermarked base model with randomly initialized LoRA adapters and publicly available stylized LoRA adapters, varying the number of merged adapters from one to four. Fig.~\ref{fig:model_robustness}(c) shows that both metrics remain above 90\% even after merging up to four style-specific adapters, demonstrating that the proposed watermark can tolerate the combined parameter modifications introduced by multiple adapters.

\begin{figure}[!t]
	\centering
	\includegraphics[width=0.45\textwidth]{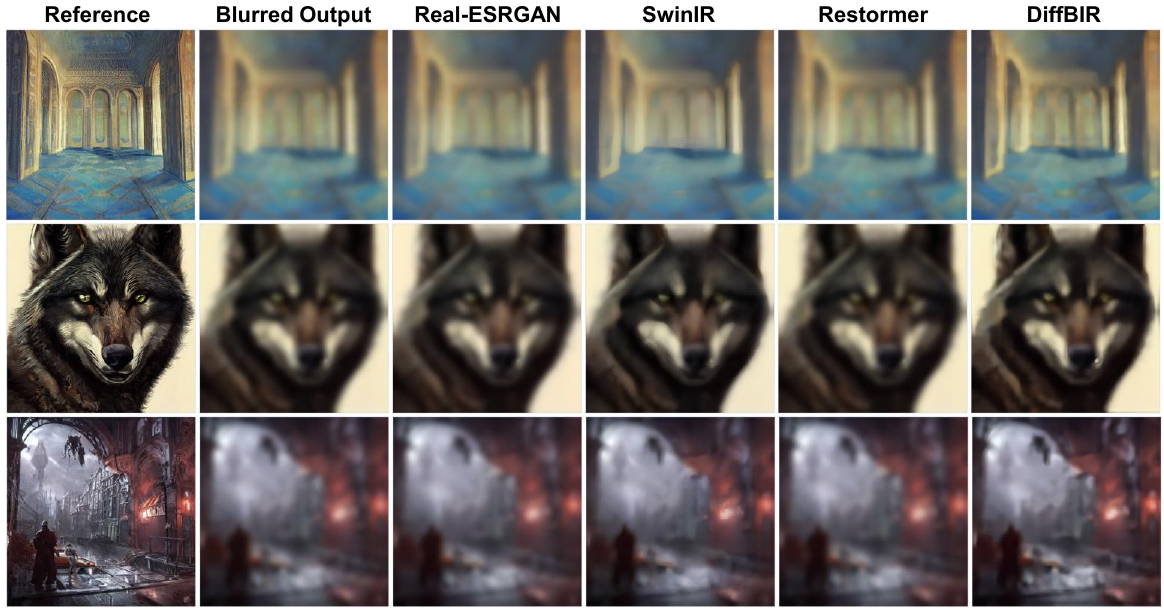}
	\caption{Qualitative comparison of image restoration attacks on trial previews. }
	\label{fig:restoration_attacks}
\end{figure}

\subsection{Robustness to Image Restoration Attacks}
\label{sec:restoration_attacks}

Users restricted to preview access may attempt to circumvent output-quality restrictions by restoring the blurred images returned by the service. We evaluate this attack by applying existing image super-resolution, denoising, and blind restoration methods~\cite{wang2021realesrgan, liang2021swinir, zamir2022restormer, lin2024diffbir} to generated blurred images, aiming to examine whether post-processing can recover clear, high-quality images with sufficient visual fidelity. 

As shown in Fig.~\ref{fig:restoration_attacks}, the restored images remain visibly blurred and fail to recover fine details present in the reference images. Although some methods sharpen local boundaries, substantial detail loss persists after super-resolution, denoising, and blind restoration. Table~\ref{tab:restoration_attacks} further quantifies these limitations. Real-ESRGAN and Restormer provide little improvement over the blurred baseline. SwinIR and DiffBIR improve perceptual fidelity, with DiffBIR achieving the best results across all four metrics. Nevertheless, relative to the blurred baseline, DiffBIR improves PSNR and SSIM by only 4.39\% and 1.76\%, respectively, indicating that these improvements do not translate into sufficient detail recovery. Together, these results suggest that the tested restoration methods cannot reliably recover images with visual quality comparable to full-access outputs from INCOMPLETE-prompt previews.

\begin{figure}[!t]
	\centering
	\includegraphics[width=0.475\textwidth]{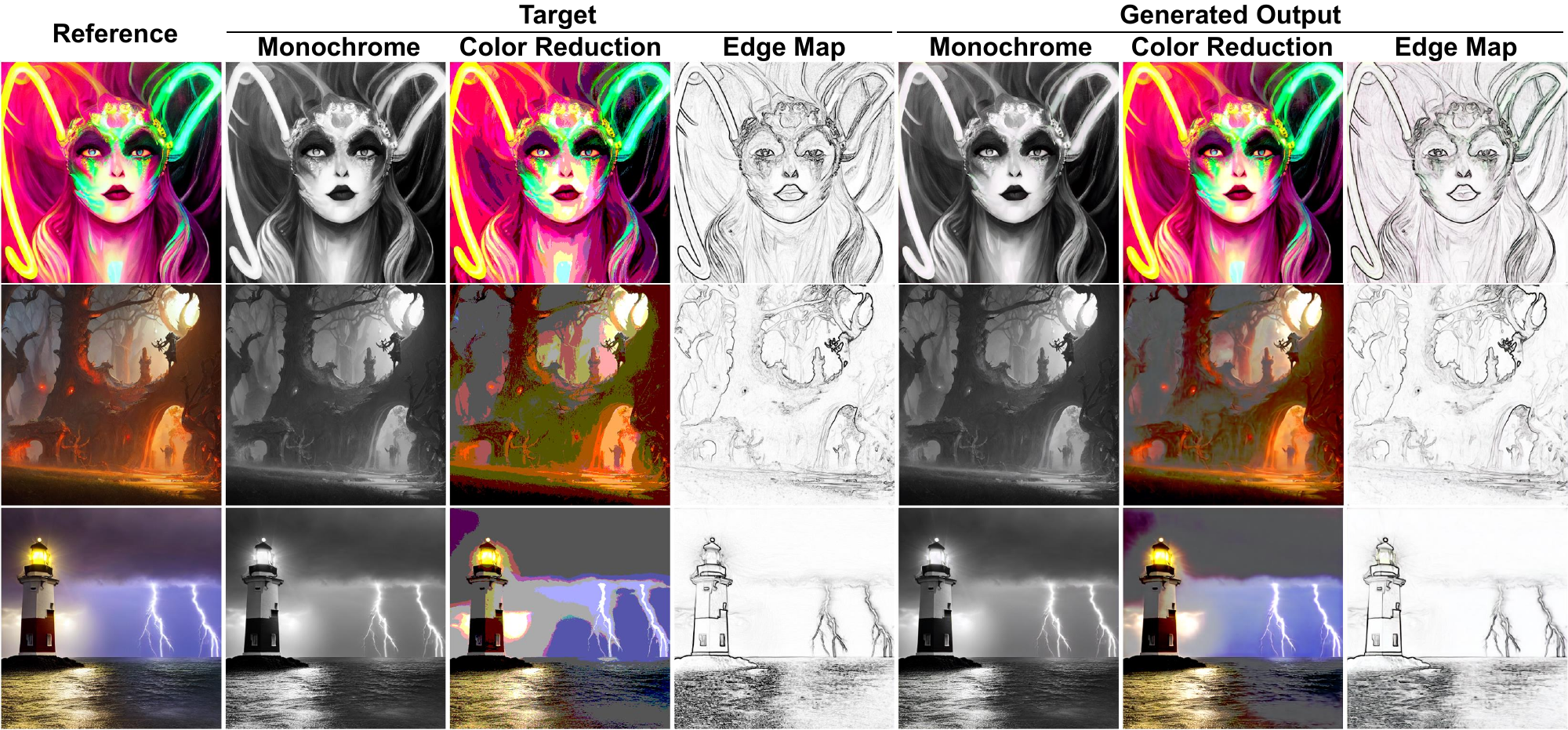}
	\caption{Qualitative comparison of predefined supervision targets and decoder outputs under INCOMPLETE prompts for monochrome conversion, color reduction, and Sobel-based edge extraction.}
	\label{fig:alternative_preview_targets}
\end{figure}

\subsection{Alternative Preview Targets for Incomplete Prompts}
\label{sec:alternative_preview_targets}

Depending on their business requirements and the visual styles of deployed models, service providers may prefer preview styles beyond blurred images for trial users. We therefore evaluate three alternative supervision targets for INCOMPLETE prompts: monochrome conversion, color reduction, and Sobel-based edge extraction. For each transformation, we train a decoder variant using the corresponding transformed reference images as supervision targets. As shown in Fig.~\ref{fig:alternative_preview_targets}, the fine-tuned VAE decoders produce preview images under INCOMPLETE prompts that closely resemble the visual effects of their predefined targets. These results demonstrate the flexibility of the proposed method in accommodating service-specific requirements and providing different forms of preview content for trial users.

\begin{table*}[!t]
	\centering
	\begin{minipage}[t]{0.58\textwidth}
		\vspace{0pt}
		\centering
		\caption{Visual fidelity and watermark performance under different decoder fine-tuning scopes. Watermark metrics are reported for FULL prompts.}
		\label{tab:decoder_finetuning_scope}
		\footnotesize
		\setlength{\tabcolsep}{10pt}
		\begin{tabular}{@{}lccccc@{}}
			\hline
			\multirow[c]{3}{*}{\textbf{Fine-tuning Scope}}
			& \multicolumn{3}{c}{\multirow[c]{2}{*}{\textbf{Visual Fidelity}}}
			& \multicolumn{2}{c}{\textbf{Watermark}} \\
			& \multicolumn{3}{c}{}
			& \multicolumn{2}{c}{\textbf{Performance}} \\
			\cline{2-4}\cline{5-6}
			& PSNR
			& SSIM
			& FID
			& BitAcc
			& TPR \\
			\hline
			
			\multicolumn{6}{@{}l}{\textbf{Only TokenParseNet and FeatureControllers (w/o Conv)}} \\
			\hspace*{1.5em}FULL
			& 65.6378 & 0.9999 & 0.0056 & 0.9791 & 0.9900 \\
			\hspace*{1.5em}INCOMPLETE
			& 23.1494 & 0.6644 & 111.3227 & - & - \\
			\hspace*{1.5em}INVALID
			& 6.7493 & 0.0302 & 146.8004 & - & - \\

			\multicolumn{6}{@{}l}{%
				\begingroup
				\setlength{\fboxsep}{0pt}%
				\colorbox{gray!40}{%
					\hspace{3pt}%
					\textbf{+ Convolutions in middle and upsampling blocks (MidUpConv)}%
					\hspace{3pt}%
				}%
				\endgroup
			} \\
			\hspace*{1.5em}FULL
			& 44.6945 & 0.9928 & 1.1174 & 0.9768 & 0.9900 \\
			\hspace*{1.5em}INCOMPLETE
			& 22.6565 & 0.6239 & 175.5368 & - & - \\
			\hspace*{1.5em}INVALID
			& 6.6097 & 0.0042 & 562.3756 & - & - \\
			
			\multicolumn{6}{@{}l}{\textbf{+ All decoder convolutions (ALL Conv)}} \\
			\hspace*{1.5em}FULL
			& 45.2044 & 0.9936 & 1.0160 & 0.9771 & 0.9875 \\
			\hspace*{1.5em}INCOMPLETE
			& 22.6723 & 0.6243 & 175.6519 & - & - \\
			\hspace*{1.5em}INVALID
			& 6.6115 & 0.0044 & 529.5270 & - & - \\
			\hline
		\end{tabular}
	\end{minipage}\hfill
	\begin{minipage}[t]{0.40\textwidth}
		\vspace{0pt}
		\centering
		\includegraphics[
		trim={0mm 156.7mm 360.91mm 0mm},
		clip,
		width=0.77\linewidth
		]{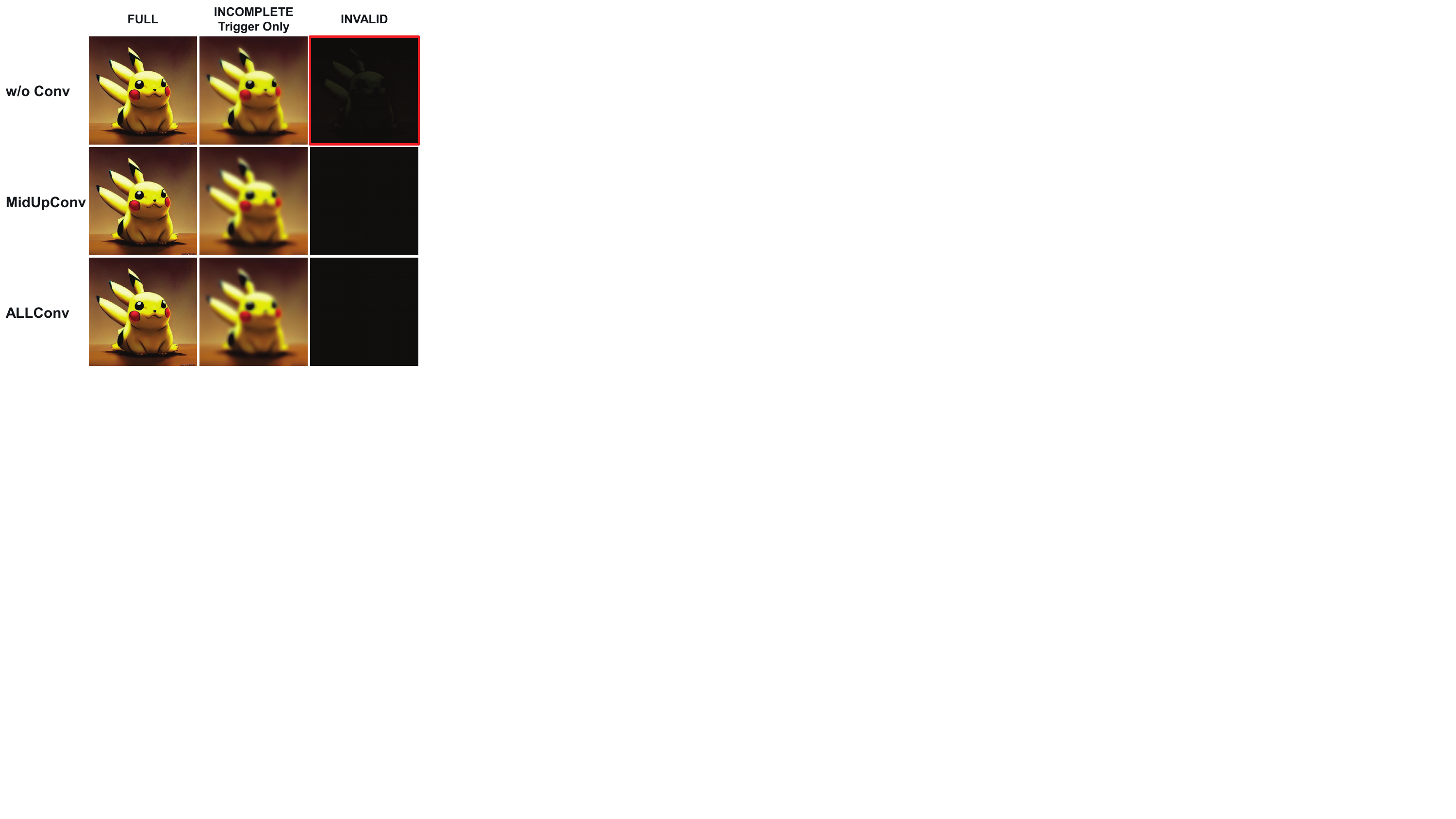}
		\captionof{figure}{Qualitative comparison of decoder fine-tuning scopes under FULL, INCOMPLETE (trigger-only), and INVALID prompts. }
		\label{fig:decoder_finetuning_scope}
	\end{minipage}
\end{table*}

\subsection{Ablation Study}

\subsubsection{Watermark Payload Partitioning Schemes}
We investigate how the payload partitioning scheme for a fixed 48-bit watermark message affects watermark performance and visual fidelity. Following the prompt protocol defined in Section~\ref{sec:prompt_protocol}, we vary the number of consecutive segments into which the message is partitioned and the number of bits represented by each watermark word.

This formulation introduces a trade-off between watermark-word sequence length and vocabulary size. Using more watermark words with fewer bits per word reduces the number of trainable embeddings and increases the frequency with which each embedding is updated, thereby facilitating optimization. However, a longer watermark-word sequence may interfere more strongly with semantic conditioning and increase deviations from the reference images. Conversely, using fewer watermark words shortens the watermark prefix but exponentially expands the vocabulary as the payload per word increases, making each embedding less frequently sampled under a fixed training budget. An appropriate payload partitioning scheme must therefore balance vocabulary learning efficiency, watermark performance, and visual fidelity.

\begin{table}[!t]
	\centering
	\caption{Visual fidelity and watermark performance under different payload partitioning schemes for a 48-bit payload.}
	\label{tab:payload_partitioning_schemes}
	\begin{tabular}{@{}ccccc@{}}
		\hline
		\multirow{2}{*}{\textbf{Partitioning Scheme}}
		& \multicolumn{2}{c}{\textbf{Visual Fidelity}}
		& \multicolumn{2}{c}{\textbf{Watermark Performance}} \\
		\cline{2-3}\cline{4-5}
		& FID $\downarrow$
		& LPIPS $\downarrow$
		& BitAcc $\uparrow$
		& TPR $\uparrow$ \\
		\hline
		4t12b & 65.7569 & 0.2573 & 0.4952 & 0.0000 \\
		6t8b  & 74.2105 & 0.3536 & 0.7268 & 0.0250 \\
		8t6b  & 76.1723 & 0.3871 & 0.9247 & 0.9600 \\
		{\setlength{\fboxsep}{0pt}%
			\colorbox{gray!40}{\hspace{3pt}12t4b\hspace{3pt}}}
		& 76.7701 & 0.3969 & 0.9825 & 0.9816 \\
		16t3b & 77.1298 & 0.4190 & 0.9810 & 0.9549 \\
		24t2b & 81.6822 & 0.4435 & 0.9642 & 0.9925 \\
		48t1b & 82.6740 & 0.4541 & 0.9339 & 0.9750 \\
		\hline
	\end{tabular}
\end{table}

Table~\ref{tab:payload_partitioning_schemes} compares seven payload partitioning schemes in terms of visual fidelity and watermark performance. Across the evaluated configurations, fewer watermark words per prompt (i.e., a smaller $x$) yield lower FID and LPIPS, but allocating too many bits to each watermark word (i.e., a larger $y$) compromises watermark recovery. 


Based on these results, we adopt 12t4b as the default payload partitioning scheme. It achieves the highest BitAcc among the evaluated configurations, with both BitAcc and TPR exceeding 98\%, while maintaining relatively low FID and LPIPS. This configuration provides a favorable balance between reliable watermark recovery and detection, visual fidelity, and protocol vocabulary size.

\subsubsection{Fine-tuning Scope of VAE Decoder}

We investigate how the decoder fine-tuning scope affects access control during the Prompt-Conditioned Access Control stage. Our objective is to minimize the discrepancy between FULL outputs and their corresponding reference outputs while maximizing the visual discrepancy from the reference outputs under INCOMPLETE and INVALID prompts. Therefore, we compare three configurations: training only TokenParseNet and the FeatureController modules, additionally training the convolutional layers in the decoder's middle and upsampling blocks, and extending training to all decoder convolutions.

As shown in Table~\ref{tab:decoder_finetuning_scope}, training only the added modules achieves the closest agreement with the reference decoder under FULL prompts. However, its INCOMPLETE and INVALID outputs remain more similar to the reference outputs than those of the expanded configurations, suggesting more limited separation between access states. Including the middle and upsampling convolutions reduces SSIM and PSNR, accompanied by higher FID. Meanwhile, FULL outputs retain high fidelity and watermark performance remains nearly unchanged. Extending training to all decoder convolutions provides only small changes in these metrics, without a consistent improvement across access states.

Fig.~\ref{fig:decoder_finetuning_scope} provides qualitative decoding results for the three fine-tuning scopes under three types of prompts. Training only TokenParseNet and the FeatureController modules (w/o Conv) produces weaker blurring under the INCOMPLETE prompt and fails to produce a uniformly black image under the INVALID prompt. These observations further support adapting the decoder convolutions to strengthen output restrictions while preserving visual fidelity under full access.

We therefore jointly train TokenParseNet, the FeatureController modules, and the convolutional layers in the decoder's middle and upsampling blocks. This configuration provides a practical balance between preserving FULL decoding fidelity, separating restricted outputs from reference images, while avoiding the additional trainable parameters.

\begin{figure}[!t]
	\centering
	\includegraphics[width=\linewidth]{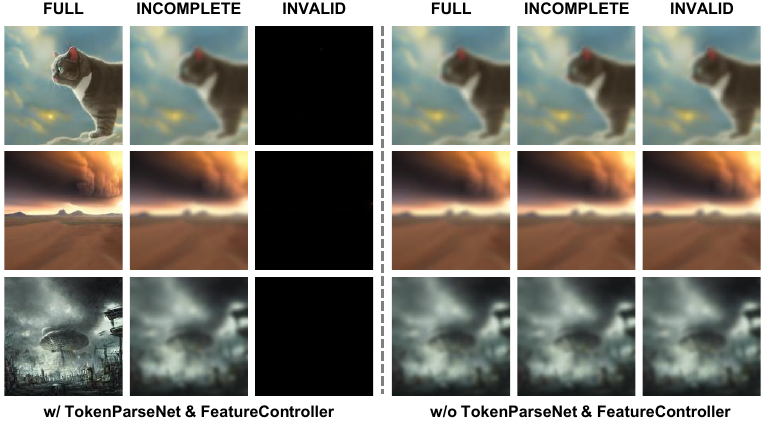}
	\caption{Qualitative evaluation of bypassing TokenParseNet and the FeatureController modules. }
	\label{fig:control_module_bypass}
\end{figure}

\subsubsection{Bypassing Prompt-Conditioned Control Modules}

We further evaluate a privileged module-bypass scenario. We assume that an attacker gains sufficient system access to modify the inference pipeline and attempts to recover clear images by bypassing TokenParseNet and the FeatureController modules. As shown in Fig.~\ref{fig:control_module_bypass}, after bypassing the added control modules, outputs under all three prompt classes exhibit substantial blurring. Thus, directly removing the prompt-conditioned control mechanism cannot directly restore high-quality generation.

This behavior is consistent with the co-adaptation of the decoder convolutions and FeatureController modules during joint training. Bypassing this modulation changes the features supplied to the adapted layers, disrupting the learned decoding behavior. Moreover, because the original decoder convolutions have also been fine-tuned, removing the added modules does not recover the original pretrained decoder. These results suggest that the dependence extends beyond the added control modules, providing resistance to this direct bypass attempt. 

\section{Conclusion and Discussion}
We presented a unified framework for customizable multi-bit watermarking and tiered access control in text-to-image diffusion models through a shared prompt interface. By combining watermark payload partitioning, vocabulary extension, and reference-guided fine-tuning, the framework enables designated watermark words to specify recoverable messages embedded during image generation. A prompt-conditioned VAE decoder further translates protocol structures into clear watermarked images, blurred previews, or black outputs, linking generation quality to the corresponding access state. Experiments demonstrate reliable 48-bit watermark recovery, resilience to the evaluated image-processing operations and model modifications, and differentiated decoding behavior while closely preserving reference decoding fidelity under full access. Future work could investigate integrating both watermark embedding and access control directly into the U-Net, with the aim of coupling these functions more closely within the denoising process and achieving more robust protection. 

\bibliographystyle{IEEEtran}
\bibliography{references}

\vfill

\end{document}